\documentclass[apj]{emulateapj}

\usepackage{mathptmx}
\usepackage{color}
\usepackage{subfigure}

\journalinfo{Draft version \today}
\slugcomment{ApJ in press}
\shorttitle{A Dynamical Study of Nova Muscae 1991}
\shortauthors{WU ET AL.}

\def\simgt{\lower 2pt \hbox{$\, \buildrel {\scriptstyle >}\over {\scriptstyle \sim}\,$}}
\def\simlt{\lower 2pt \hbox{$\, \buildrel {\scriptstyle <}\over {\scriptstyle \sim}\,$}}

\def\ha{H$\alpha$}

\def\vsini{$v\sin i$}

\def\xray{\hbox{X-ray}}

\def\nmus{NovaMus}

\begin{document}
\renewcommand*{\thefootnote}{\fnsymbol{footnote}}

\title{A Dynamical Study of the Black Hole X-ray Binary Nova Muscae
  1991\footnotemark[1]}



\author{\rm Jianfeng~Wu\altaffilmark{2},
Jerome~A.~Orosz\altaffilmark{3},
Jeffrey~E.~McClintock\altaffilmark{2},
Danny~Steeghs\altaffilmark{4,2},
Pen\'{e}lope~Longa-Pe\~{n}a\altaffilmark{4},
Paul~J.~Callanan\altaffilmark{5},
Lijun~Gou\altaffilmark{6},
Luis~C.~Ho\altaffilmark{7,8},
Peter~G.~Jonker\altaffilmark{9,10,2},
Mark~T.~Reynolds\altaffilmark{11},
Manuel~A.~P.~Torres\altaffilmark{9,10}
}

\altaffiltext{1}
     {This paper includes data gathered with the 6.5 meter Magellan
     Telescopes located at Las Campanas Observatory, Chile.}
\altaffiltext{2}
     {Harvard-Smithsonian Center for Astrophysics, 60 Garden Street,
     Cambridge, MA 02138, USA} 
\altaffiltext{3}
     {Department of Astronomy, San Diego State University, 5500
     Campanile Drive, San Diego, CA 92182, USA} 
\altaffiltext{4}
    {Department of Physics, University of Warwick, Coventry, CV4~7AL,
     UK} 
\altaffiltext{5}
    {Department of Physics, University College Cork, Cork, Ireland} 
\altaffiltext{6}
    {National Astronomical Observatories, Chinese Academy of Sciences,
     Beijing 100012, China} 
\altaffiltext{7}
    {Kavli Institute for Astronomy and Astrophysics, Peking
    University, Beijing 100871, China}
\altaffiltext{8}
    {Department of Astronomy, School of Physics, Peking University,
     Beijing 100871, China} 
\altaffiltext{9}
    {SRON, Netherlands Institute for Space Research, Sorbonnelaan 2,
     3584 CA, Utrecht, The Netherlands}
\altaffiltext{10}
    {Department of Astrophysics/IMAPP, Radboud University Nijmegen,
     Heyendaalseweg 135, 6525 AJ, Nijmegen, The Netherlands}
\altaffiltext{11}
    {Department of Astronomy, University of Michigan, 1085
     S. University Avenue, Ann Arbor, MI 48109, USA} 
\email{jianfeng.wu@cfa.harvard.edu}


\begin{abstract}

  We present a dynamical study of the Galactic black hole binary system
  Nova Muscae 1991 (GS/GRS 1124$-$683). We utilize 72 high resolution
  Magellan Echellette (MagE) spectra and 72 strictly simultaneous
  $V$-band photometric observations; the simultaneity is a unique and
  crucial feature of this dynamical study. The data were taken on two
  consecutive nights and cover the full 10.4-hour orbital cycle.  The
  radial velocities of the secondary star are determined by
  cross-correlating the object spectra with the best-match template
  spectrum obtained using the same instrument configuration.  Based on
  our independent analysis of five orders of the echellette spectrum, the
  semi-amplitude of the radial velocity of the secondary is measured to
  be $K_{\rm 2} = 406.8\pm2.7$~km~s$^{-1}$, which is consistent with
  previous work, while the uncertainty is reduced by a factor of 3. The
  corresponding mass function is $f(M)=3.02\pm0.06\ M_\odot$. We have
  also obtained an accurate measurement of the rotational broadening of
  the stellar absorption lines ($v\sin i=85.0\pm2.6$~km~s$^{-1}$) and
  hence the mass ratio of the system $q=0.079\pm0.007$. Finally, we have
  measured the spectrum of the non-stellar component of emission that
  veils the spectrum of the secondary. In a future paper, we will use
  our veiling-corrected spectrum of the secondary and accurate values of
  $K_{\rm 2}$ and $q$ to model multi-color light curves and determine
  the systemic inclination and the mass of the black hole.

\end{abstract}


\renewcommand*{\thefootnote}{\arabic{footnote}}
\setcounter{footnote}{0}
\section{Introduction}\label{intro}

Mass is the fundamental parameter for a black hole. According to the
No-Hair Theorem (e.g., Israel 1967; Hawking 1971), mass and spin
together provide a complete description of an astrophysical black
hole. An accurate measurement of mass is a prerequisite for measuring
black hole spin via the continuum-fitting method (see McClintock
et~al. 2014 for a review).  Stellar-mass black holes in the Milky Way
and neighboring galaxies are discovered in \xray\ binary systems, some
of which are persistent \xray\ sources, while others are \xray\
transients which have gone through one or more \xray\ outbursts
(Remillard \& McClintock 2006). About two dozen of these black holes
have been dynamically confirmed to have masses in the range $M=5$--$30\
M_\odot$.


X-ray Nova Muscae 1991 (hereafter \nmus), with an orbital period of
10.4~hr, is one of about a dozen black-hole transient systems that are
distinguished by their short orbital periods, $P < 12$~hr. (For a sketch
to scale of nine of these systems, including the prototype A0620$-$00,
see Figure 1 in McClintock et al.\ 2014.)  \nmus\ was discovered during
its 1991 outburst independently by the {\it Ginga} (Makino et~al. 1991)
and {\it GRANAT} (Lund \& Brandt 1991) missions.  Eight days after
discovery, it reached a peak X-ray intensity of 8 Crab in the 1--6 keV
band (Ebisawa et~al. 1994).  Over the course of eight months, the source
passed through all the canonical X-ray states before returning to its
quiescent state (Ebisawa et al. 1994; Remillard \& McClintock 2006).
Esin et~al. (1997) used the 1991 outburst data for \nmus\ to develop a
model of the states that combines the standard model of a thin accretion
disk (Shakura \& Sunyaev 1973) and the advection-dominated accretion
flow (ADAF) model (Narayan \& Yi 1994).


Interestingly, hard X-ray observations of \nmus\ made at the peak of the
outburst revealed evidence for positron-electron annihilation in the
form of a relatively narrow and variable emission line near 500 keV
(Sunyaev et al. 1992; Goldwurm et al. 1992).  Kaiser \& Hannikainen
(2002) argued that the line could be generated by annihilation in the
bipolar outflow of the system. Mart{\'{\i}}n et~al. (1996) suggested an
alternative mechanism, namely, Li production near the black hole that
gives rise to a 476 keV nuclear line; their supporting evidence is their
detection of a relatively strong lithium absorption line $\lambda6708$
in the optical spectrum.

Soon after its discovery, \nmus\ was considered to be a strong candidate
for a black hole binary based on its X-ray spectral properties and light
curve, and its similarities to the prototype black-hole X-ray transient
A0620$-$00 (McClintock \& Remillard 1986).  Following the discovery of
the optical counterpart (Della Valle et~al. 1991), the black hole nature
of the primary was established in the customary way by measuring the
mass function:
\begin{equation}
f(M) \equiv \frac{{P}K_{2}^{3}}{2\pi G} = \frac{M\sin^3i}{(1+q)^{2}},
\end{equation} 
where $K_2$ is the semi-amplitude of the velocity curve of the
secondary, $i$ is the orbital inclination angle of the binary and $q$ is
the ratio of the companion star mass $M_2$ to that of the compact
primary $M$.  Three dynamical studies performed in quiescence found a
value for the mass function (the minimum mass of the compact primary) of
$\approx 3M_\odot$ (Remillard et~al. 1992; Orosz et~al. 1996; Casares
et~al. 1997), which is widely considered to be the maximum stable mass
of a neutron star (Kalogera \& Baym 1996).

Orosz et~al. (1994) obtained a quite uncertain estimate of the mass
ratio $q \equiv M_2/M$ by measuring the radial velocity of the H$\alpha$
emission line.  Casares et~al. (1997) made a conventional and direct
measurement of the mass ratio by measuring the rotational broadening of
the photospheric absorption lines of the secondary; they obtained a
result that is consistent with that of Orosz et~al. (1994).

The inclination angle $i$ is usually estimated by modeling the
ellipsoidal modulation of the light curves of the secondary star after a
system has returned to its quiescent state.  This modeling is
complicated in the case of the short-period systems because the stellar
component of light is significantly contaminated by the emission from
the accretion disk; this so-called ``disk veiling,'' can be quite
significant ($\gtrsim50$\% in the $V$-band) and variable in ``active''
quiescent states (Cantrell et al. 2010).  Orosz et~al. (1996) estimated
a disk contribution at $\sim5000$~\AA\ that could be as high as
50\%. Meantime, using different data, Casares et~al. (1997) estimated a
disk contribution at H$\alpha$~$\lambda$6563 of 15\%, suggesting that
disk veiling may be less significant at longer wavelengths.  Based on
this notion, Gelino et~al. (2001) modeled near-infrared ($J$ and $K$)
light curves of \nmus\ assuming that the disk contribution could be
ignored in these bands, and they thereby estimated the systemic
inclination to be $i=54^\circ \pm1.5^\circ$; their estimate of black
hole mass (which also depends on the mass ratio $q$) is $M = 6.95\pm0.6\
M_\odot$.

These estimates of $i$ and $M$ are quite uncertain because they rest on
the assumption that the effects of veiling are negligible in the
near-infrared, while there is considerable evidence that it may not be
(e.g., Reynolds et~al. 2007, 2008; Kreidberg et al. 2012).  In this
paper, we obtain the first detailed and secure measurement of the
spectrum of the non-stellar component of light that veils the stellar
spectrum.  While this achievement is in part due to the quality and
quantity of our data, it is based primarily on the strict simultaneity
of our spectroscopic and photometric observations.  The simultaneity is
crucial because of the aperiodic variability of the non-stellar
component.  Our two other principal results are a greatly refined
measurement of the $K$-velocity and the first robust measurement of the
mass ratio $q$, which differs significantly from the earlier estimates
mentioned above.

This paper is structured as follows. Observations and data reduction are
described in \S\ref{obs}. We derive the mass function via a radial
velocity analysis in \S\ref{rvc}.  The mass ratio is determined via
rotational broadening in \S\ref{vsini}. The disk veiling is also
measured in this section. The light curve of \nmus\ 
is presented in \S\ref{lc}. Finally, we discuss our results in
\S\ref{discuss}.  The light curve modeling and the determination of the
inclination and black hole mass will be presented in a subsequent paper.


\begin{figure*}[t]
    \centering
    \includegraphics[width=5.5in]{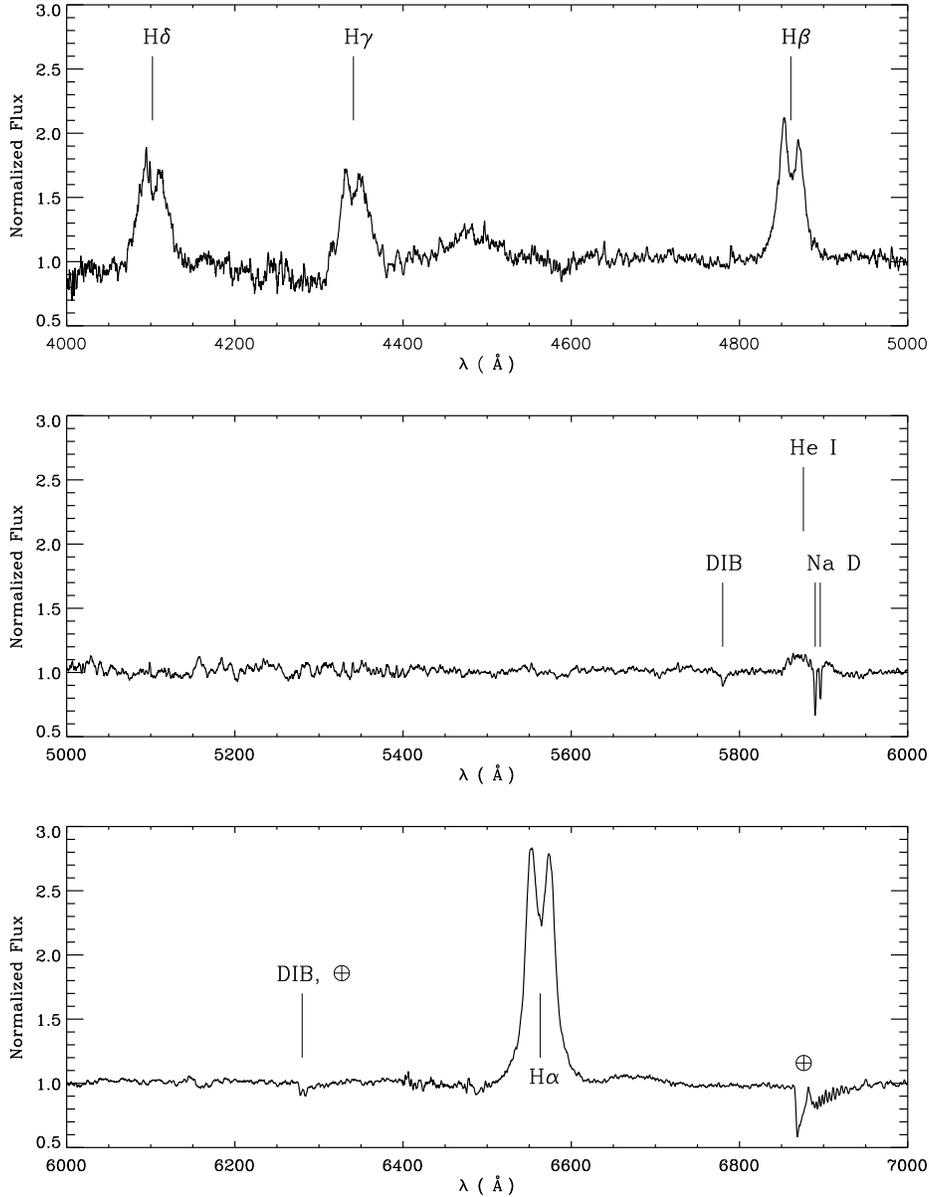}
    \caption{\footnotesize{The mean Magellan/MagE spectrum of \nmus\ in
    the observer's rest frame, covering MagE orders \#9--14
    (4000--7000~\AA). Broad and double-peaked Balmer lines (H$\alpha$,
    H$\beta$, H$\gamma$, and H$\delta$) are prominent in the spectrum. He~{\sc i}
    $\lambda$5875 emission is also evident. The weak, broad emission
    features redward of H$\gamma$ and H$\alpha$ are likely to be He~{\sc i}
    $\lambda$4471 and $\lambda$6678, respectively. The Na~D
    $\lambda\lambda$5890--5896 doublet, DIB features at $\lambda$5780 and
    $\lambda$6280 (which also has a strong telluric component), as
    well as the telluric feature at $\lambda$6860 are also
    labeled. The \nmus\ spectrum covering the full MagE wavelength range
    ($\sim3000$--$10000$~\AA) is available in the online journal in
    electronic form.}  
    \label{spec_fig}}
\end{figure*}%

\section{Observation and Data Reduction}\label{obs}

\subsection{Magellan Echellette Spectroscopy}\label{obs:spec}

The spectroscopic data were obtained using the Magellan Echellette
spectrograph (MagE; Marshall et~al. 2008) mounted on the 6.5-meter
Magellan/Clay Telescope at the Las Campanas Observatory (LCO) in Chile.
All the data were taken on the nights of 2009 April 25--26 UT. A total of
72 spectra were obtained, each with an integration time of 600~s.
During the same two nights, and with the same setup, we obtained spectra
of 38 stars, principally K dwarfs, that serve as radial velocity
templates.  The seeing was similar on both nights, with a typical value
of $\approx0.7^{\prime\prime}$ and a range of
0.5$^{\prime\prime}$--1.3$^{\prime\prime}$.  Each MagE spectrum consists
of 15 orders (corresponding to sequence numbers \#6--20).  The full
wavelength coverage of the instrument is 3100--11000~\AA, with
order~\#20 at the blue end (central wavelength $3125$~\AA) and order~\#6
at the red end (central wavelength $9700$~\AA).  The spatial scale is
$\approx0.3^{\prime\prime}$/pixel.  The spectral
dispersion increases from 0.2~\AA\ per pixel for order~\#20 to
0.7~\AA\ per pixel for order~\#6, while the velocity dispersion is 
22~km~s$^{-1}$ per pixel for all orders. For our chosen
0.85$^{\prime\prime}$ slit, the spectral resolution is
$\approx1$~\AA\ at 5000~\AA, corresponding to $\approx60$~km~s$^{-1}$.

The MagE data were reduced using the pipeline developed by Carnegie
Observatories\footnote{See
  http://code.obs.carnegiescience.edu/mage-pipeline.}, which performs
bias correction, flat-fielding, and wavelength calibration. The bias was
corrected using the overscan region of the detectors. Three sets of
flats were utilized in the calibration: 1) order definition flats taken
with the Xe-Flash lamps; 2) blue flats to define the flat field at the
blue end of a spectrum (also taken with the Xe-Flash lamps); and 3) red
flats to define the flat field at the red end of a spectrum and to
correct the fringing, which were taken with the Qf lamps. The wavelength
calibration was performed using arc spectra obtained with Thorium-Argon
lamps. The RMS residual of the wavelength solution is $<0.05$~\AA\ for 
orders \#8--18 and $<0.1$~\AA\ for the remaining orders. The
performance of wavelength calibration was examined using sky
emission lines; the relative wavelength offset is
negligible.  

The mean MagE spectrum of \nmus\ in the observer's rest-frame is shown
in Fig.~\ref{spec_fig}, which covers spectra that span orders \#9--14
($\sim4300$--7000~\AA). The spectrum in each order was normalized by a
fifth-order polynomial. In the analysis that follows, we mainly use the
data in these orders because of their good signal-to-noise ($S/N$) ratio
and general lack of sky emission lines and telluric absorption features
(relative to redder orders).  The most prominent features in the
spectrum are the broad, double-peaked Balmer lines \ha~$\lambda$6563,
H$\beta$~$\lambda$4861 and H$\gamma$~$\lambda$4341, which are the
canonical signatures of a black-hole accretion disk.  Also present are
broad disk lines of He, notably He I $\lambda$5875.  Other features such
as Na~D $\lambda\lambda$5890--5896, diffuse interstellar bands (DIB) at
$\lambda5780$, $\lambda$6280, and telluric features redward of
$\lambda$6850 are also evident.

\subsection{Optical Photometry}\label{obs:photo} 

Photometric data were obtained during the same observing run using the
du~Pont 2.5~m telescope, which is also located at LCO.  Seventy-two
$V$-band images were acquired; by design, each 10-min exposure was
strictly simultaneous (time difference $<1$~s) with its corresponding
MagE spectroscopic observation\footnote{The only exception is the first
  image whose exposure time was only 60 seconds.}.  The typical seeing was
$\approx0.7^{\prime\prime}$ (with a range of 0.6--1.1$^{\prime\prime}$).
We used the $2048\times2048$ Tek\#5 CCD, which provided a field of view
of $8.85^{\prime}\times8.85^{\prime}$ at $0.259^{\prime\prime}$ per pixel.  
Bias correction
and flat-fielding were performed with standard {\sc iraf} tasks
(\verb+zerocombine+, \verb+flatcombine+, and \verb+ccdproc+). The
centroid of the optical counterpart was determined using the CIAO
routine \verb+wavdetect+ (Freeman et~al. 2002), and the quality of the
centroiding was confirmed visually.  Aperture photometry was performed
using the IDL procedure \verb+aper+. The radius of the aperture was
fixed to eight pixels ($2.1^{\prime\prime}$), which is the largest
aperture that securely excludes significant contamination of \nmus\ by a
faint field star. Choosing a smaller aperture (e.g., $\sim5$~pixels) 
would provide very similar results. 
We first produced a light curve of \nmus\ relative to
nearby field stars and then derived an absolute calibration relative to
a pair of nearby ($24^{\prime\prime}$ and $41^{\prime\prime}$) standard
stars. Our estimate of the uncertainty in the zero-point is
0.05~mag.

\begin{center}
\begin{deluxetable*}{cccccccl}
\tabletypesize{\footnotesize}
\tablecaption{Radial Velocity Analysis Results\label{rvc_table}}
\tablewidth{0pt}
\tablehead{ \colhead{Order \#} & \colhead{$K_2$} &
  \colhead{$\gamma$} & \colhead{$T_0-2454900$} & \colhead{No. of} &
  \colhead{$\chi^2/\nu$} & \colhead{$\lambda$ Coverage}
&           \colhead{Masked Features} \\
\colhead{} & \colhead{(km~s$^{-1}$)} &
  \colhead{ (km~s$^{-1}$)} & \colhead{(d)} & \colhead{Object Spectra} &
  \colhead{} & \colhead{(\AA)}
&           \colhead{} 
}
\startdata
9\tablenotemark{a} & $413.6\pm4.3$ & $17.4\pm3.1$ & $46.90550\pm0.00072$ & $39$ &
$1.55$ & 6300--7300 & H$\alpha$~$\lambda6563$, 6820--7020~\AA,
7150--7200~\AA \\
10 & $409.5\pm2.7$ & $19.4\pm2.0$ & $46.90351\pm0.00044$ & $72$ &
$0.92$ & 5680--6630 & Na~D~$\lambda\lambda5890,5896$, DIB~$\lambda$6280 \\
& & & & & & & O~{\sc i}~$\lambda$6300,
H$\alpha$~$\lambda6563$ \\
11 & $413.0\pm3.1$ & $15.2\pm2.3$ & $46.90258\pm0.00051$ & $67$ &
$0.74$ & 5170--6000 & O~{\sc i}~$\lambda5577$, 
Na~D~$\lambda\lambda5890,5896$ \\
12 & $401.2\pm2.6$ & $9.7\pm1.9$ & $46.90327\pm0.00044$ & $72$ &
$0.87$ & 4700--5500 & H$\beta$~$\lambda4861$ \\
13 & $407.0\pm4.2$ & $17.0\pm3.1$ & $46.90211\pm0.00069$ & $57$ &
$0.92$ & 4370--5100 & H$\beta$~$\lambda4861$ \\
14 & $403.9\pm3.9$ & $9.2\pm2.9$ & $46.90086\pm0.00068$ & $63$ &
$1.02$ & 4060--4730 & H$\gamma$~$\lambda4341$ \\
\hline\\
Average & $406.8\pm2.2$ & $14.2\pm2.1$ & $46.90278\pm0.00042$ & & &
\enddata
\tablecomments{The quoted uncertainties are at the $1\sigma$ level of
  confidence.}  \tablenotetext{a}{The best-fit parameters for order \#9
  were not included in calculating the average parameter values; for
  this order, the wavelength ranges 6820--7020~\AA\ and 7150--7200~\AA\
  were masked to avoid strong telluric absorption features and sky
  emission lines.}
\end{deluxetable*}
\end{center}


\begin{figure}[t]
    \centering
    \includegraphics[width=3.3in]{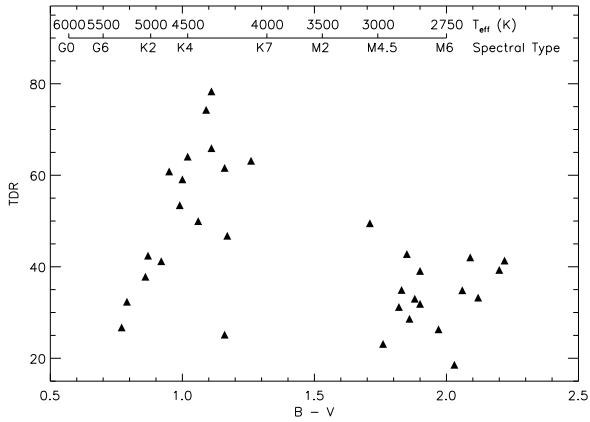}
    \caption{\footnotesize{The average TDR value of the
    cross-correlation between the spectrum of each template star and 
    \nmus\ for order \#12, plotted against the $B-V$ color of the
    star. The template star with the highest average TDR
    value is HD~170493, with $B-V=1.11$, which corresponds to a spectral type
    of $\sim$K5 with ${\rm T}_{\rm eff}\approx4400$~K, based on the
    scale of $B-V$ color vs. ${\rm T}_{\rm eff}$ vs. Spectral type,
    which is shown at the top of the figure. }
    \label{temp_fig}}
\end{figure}%

\section{Radial Velocity Analysis}\label{rvc}

The radial velocity $K_2$ of the secondary star was determined by
sequentially cross-correlating the 72 object spectra with the spectrum
of a template star.  This was done in turn for individual echellette
orders, rather than using a single merged spectrum.  The best quality
results were obtained for order \#12 (covering $\sim4700$--5500~\AA), as
expected since strong stellar absorption features are present in this
order (e.g., the Mg features at $\sim$5200~\AA\ region), and it is
relatively free of contaminating telluric and interstellar features.

\subsection{Choosing the Best Template Spectrum}\label{rvc:temp}

The cross-correlation analysis was performed using the software package
\verb+fxcor+ as implemented in the {\sc iraf} package.  The task
\verb+fxcor+ returns the parameter {\it R} defined by Tonry \& Davis
(1979; TDR hereafter) provides an estimate of $S/N$: the higher the TDR
value, the better the quality of the cross-correlation.  From among our
dozens of template spectra, we chose the one that gives the highest TDR
value, while focusing on the \nmus\ data for order \#12.  Specifically,
we: (1) selected a K-type template spectrum that gives relatively high
values of TDR and cross-correlated it against the 72 spectra of \nmus;
(2) we velocity-shifted all the \nmus\ spectra to the rest frame of the
template star and summed them; (3) we then cross-correlated this summed
spectrum against all of our template spectra in turn and selected the
template spectrum that yielded the maximum TDR value.  

The results of this analysis are summarized in Fig.~\ref{temp_fig},
which shows how TDR varies among the template stars, which we have
ordered by their $B-V$ color given in the SIMBAD database; we regard the
photometric color as a more accurate measure of the effective
temperature than the spectral type (that is given in the same
database)\footnote{We assume that the reddening of these template star
  can be ignored due to their proximity to the Sun ($\lesssim100$~pc).}.
The star at the peak of this ordered distribution and having the maximum
TDR is HD~170493 with $B-V=1.11$, corresponds to an effective
temperature $T_{\rm eff}\approx4400$~K ({\it Allen's Astrophysical
  Quantities, 4th edn.}).  We therefore choose HD~170493 as the master
template for our cross-correlation analysis. Potential systematic
effects related to this choice of template are discussed in
\S\ref{discuss:sys}.

\subsection{Fitting the Radial Velocity Curve}\label{rvc:fit}

\begin{figure*}[t]
    \centering
    \includegraphics[width=2.8in]{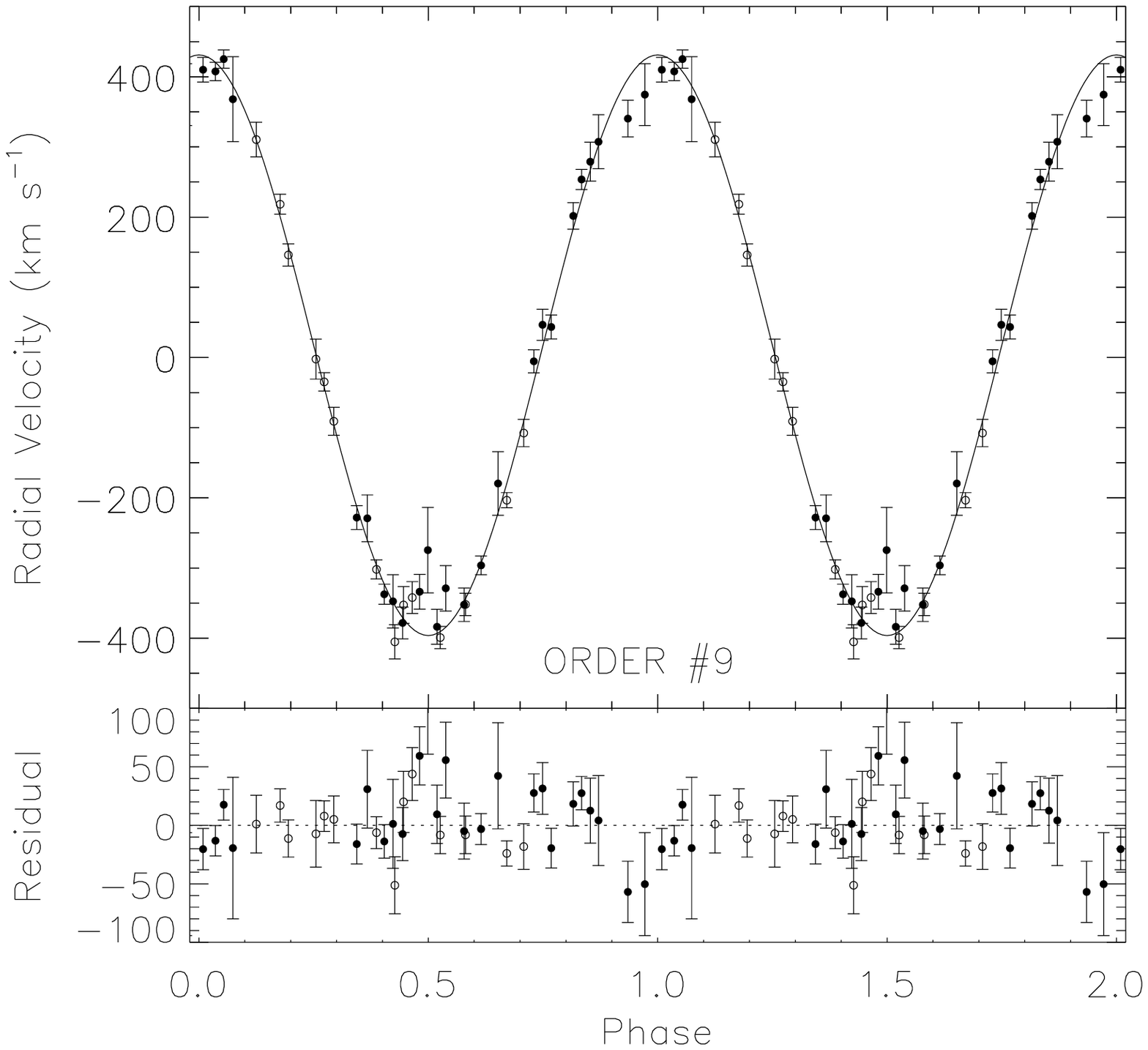}
    \includegraphics[width=2.8in]{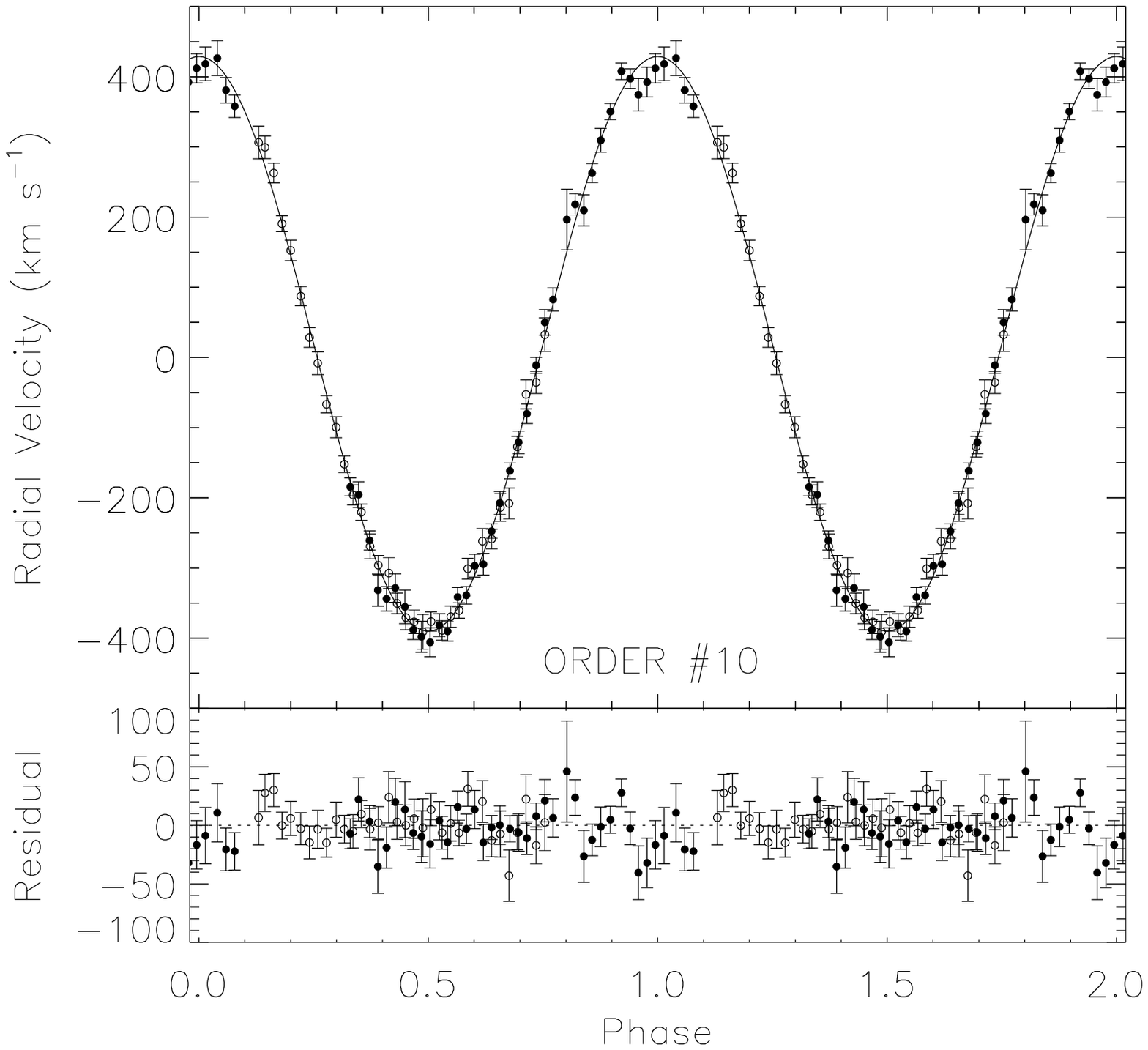}\\
    \includegraphics[width=2.8in]{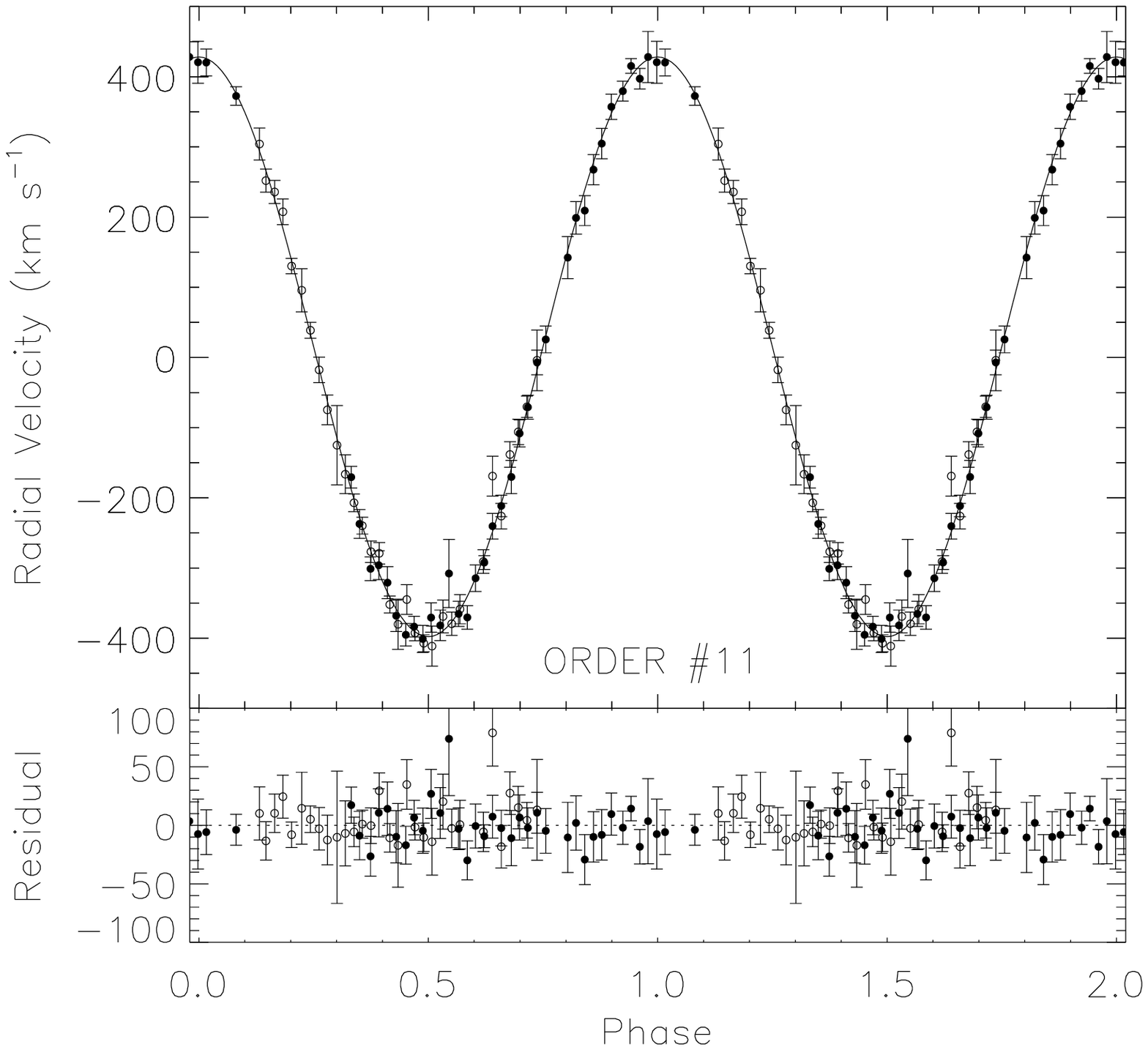}
    \includegraphics[width=2.8in]{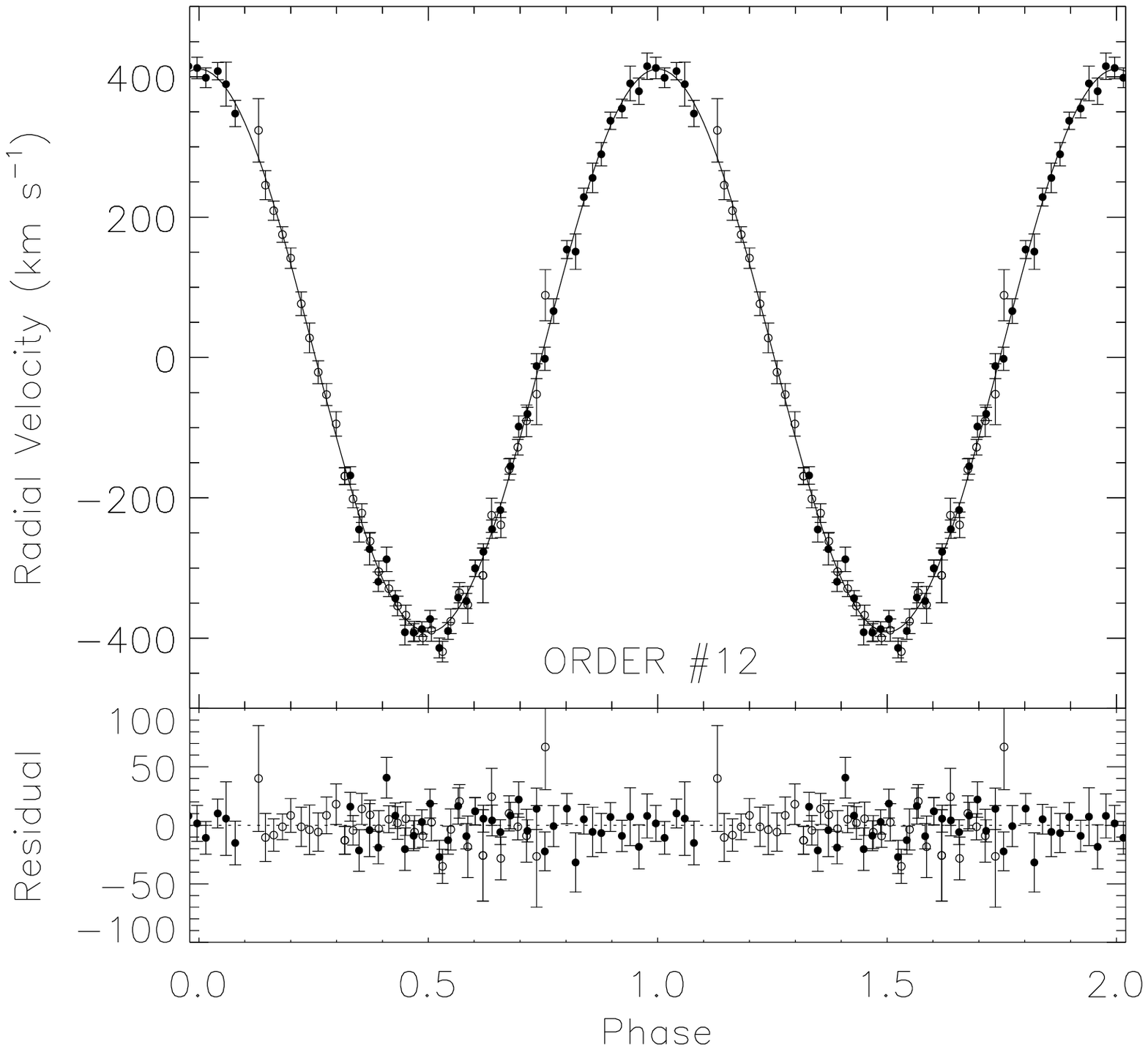}\\
    \includegraphics[width=2.8in]{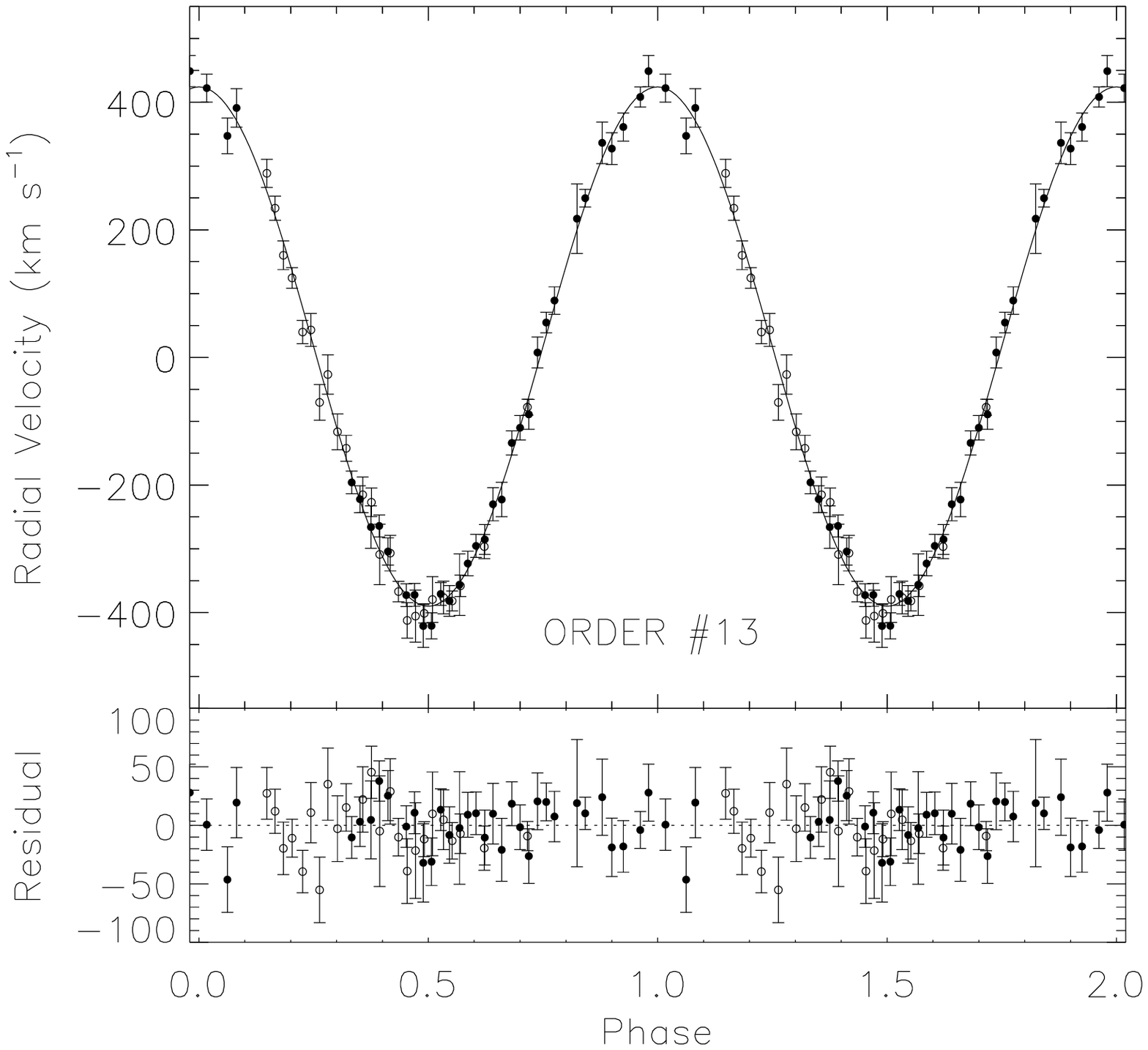}
    \includegraphics[width=2.8in]{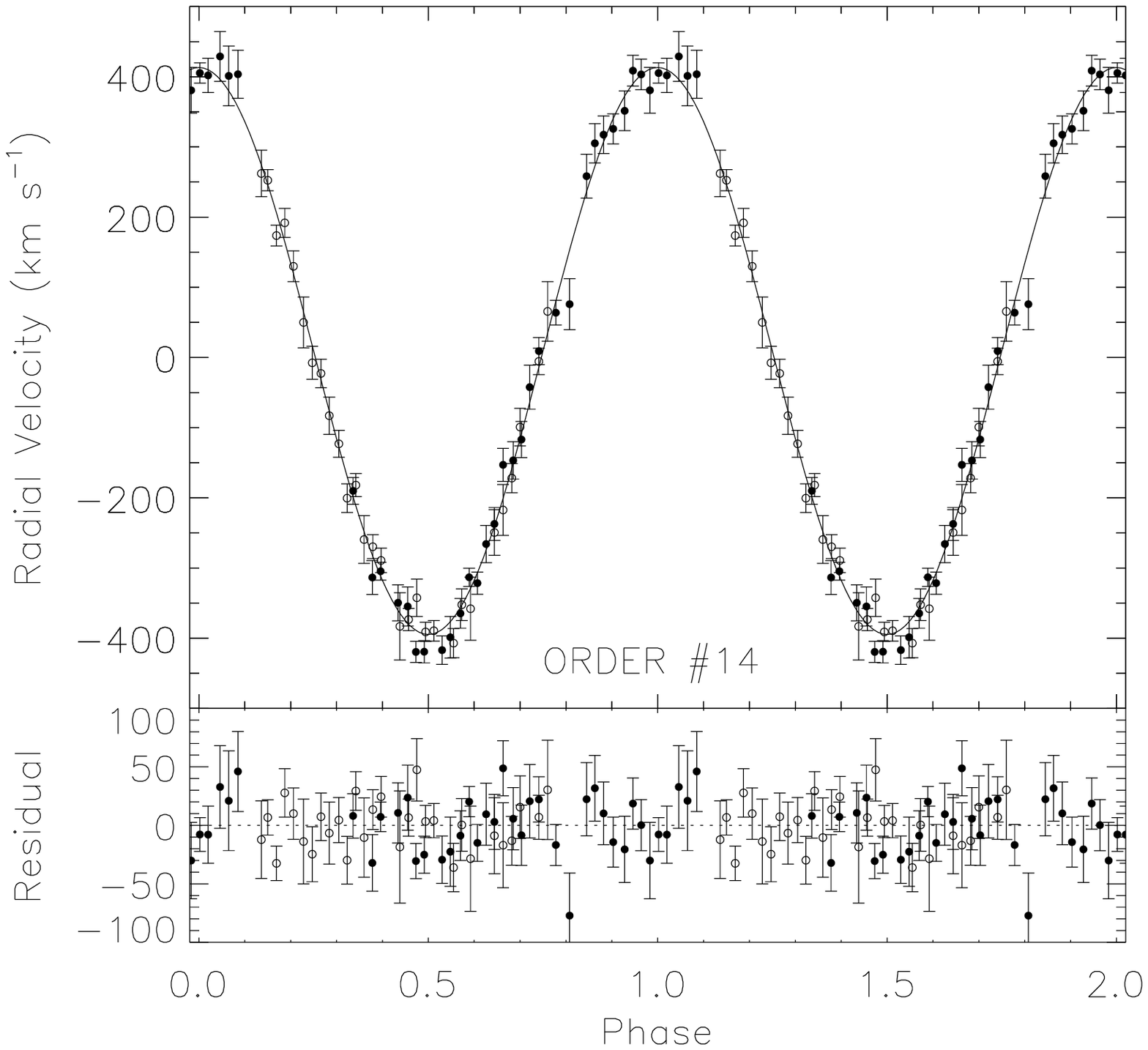}
    \caption{\footnotesize{The radial velocities of \nmus\ relative to
    the spectral template HD~170493, in MagE orders \#9--14. The
    best-fit radial velocity curves and residuals for each order are
    shown.}
             \label{rvc_fig}}
\end{figure*}%

With HD~170493 as the template, we first cross-correlated each of the 72
spectra for order \#12.  For both the object and template spectra, we
masked out the region around the H$\beta$ emission line, and also
$\sim100$~\AA\ at both ends of each order because the $S/N$ in these
regions is poor.  We furthermore required that the TDR value for a
correlation exceed 2.5 (similar to the criterion in Orosz et al. 1996).
All 72 spectra for order \#12 delivered reliable measurements of radial
velocity, as illustrated in the middle-right panel of
Fig.~\ref{rvc_fig}.  We fitted the radial velocity curve with the
following model,
\begin{eqnarray}
V(t)=\gamma+K_2\cos(2\pi\ \frac{t-T_0}{P}),
\end{eqnarray}
where $\gamma$ is the systemic velocity of in the heliocentric frame,
$t$ is the observation time in Heliocentric Julian Days (HJDs), and
$T_0$ is the time of maximum radial velocity; the radial velocity we
used for the template star HD~170493 is
$\gamma_0$~=~$-55.07\pm0.07$~km~s$^{-1}$.  We initially fixed the orbital
period to the spectroscopic value given by Orosz et~al. (1996),
$P=0.4326058(31)$~d.  Then, using our value of $T_0$ and the one in
Orosz et al. we obtained a refined value of the orbital period:
$P=0.43260249(9)$~d. This period and other best-fit parameters are
listed in Table~\ref{rvc_table}.

As a check on our value for the $K$-velocity, we performed an
independent analysis using the \verb+xcor+ procedure in the {\sc molly}
package developed by T.~Marsh\footnote{See
  http://deneb.astro.warwick.ac.uk/phsaap/software/molly/html/INDEX.html.}
in place of \verb+fxcor+ in {\sc iraf}.  We again analyzed only the data
for order \#12 and used the same template spectrum (HD~170493) as
before.  Both the object and template spectra were rebinned to the same
velocity dispersion using the \verb+vbin+ procedure, and were normalized
using a fifth-order polynomial, as required by the \verb+xcor+
procedure.  Fitting a Gaussian to the profile of the cross-correlation
function using the {\sc molly} procedure \verb+mgfit+, we obtained
values for both the $K$-velocity and its uncertainty
($K_2=402.9\pm1.3$~km~s$^{-1}$) that are consistent with those obtained
using the {\sc iraf} routines.

In order to obtain a more precise value of $K_2$, we analyzed data for
orders \#10 and \#11 (which lie to the red of order \#12) and orders
\#13 and \#14, again masking out $\sim100$~\AA\ at both ends of each
order; the wavelength ranges of the orders are given in
Table~\ref{rvc_table}.  As before, we masked out disk emission lines
(mainly Balmer lines), and we also masked out a number of strong
telluric and interstellar features, which are listed in
Table~\ref{rvc_table}).  As was the case for order \#12 discussed above,
all 72 object spectra for order \#10 yielded reliable measurements of
radial velocity (i.e., TDR $>$ 2.5); the radial velocity curve for this
order is shown in Fig.~\ref{rvc_fig}.  For the other three orders, some
object spectra were rejected because: (1) the TDR value for the
correlation was $\leq$ 2.5; (2) the fit to the cross-correlation peak
did not converge; or (3) the value obtained for the radial velocity
lies more than $\simgt3\sigma$ away from the fitted curve.\footnote{Only 
five data points in all five orders lie $\simgt3$--$4\sigma$ away from 
the fitted curve; removing these radial velocity points does not 
significantly change the fitted values of the parameters.}  The velocity
data for each order individually were fitted to the same model used for
order \#12 (Eqn.~2).  For each of the five orders, Table~\ref{rvc_table}
lists the best-fit parameters, values of reduced $\chi^2$ and the number
of useful spectra.  Using precisely the velocity uncertainties returned
by \verb+fxcor+, we obtain good fits for all five orders, with
$\chi^2/\nu$ values that are close to unity.  We note that there are
small differences among the best-fit parameters at the level of
$\sim3$--$4\sigma$ level. 

Table~\ref{orbital_table} provides a summary of the orbital
parameters. The error-weighted mean of $K_2$ among the five orders is
$406.8\pm2.2$~km~s$^{-1}$. After including our estimate of the
systematic error (see \S\ref{discuss:sys}), we obtain for the radial
velocity semi-amplitude $K_2=406.8\pm2.7$~km~s$^{-1}$. Using this value
of $K_2$, the orbital period $P$ (Table~\ref{orbital_table}) and Eqn.\
(1), we derive a precise value for the mass function of
$f(M)=3.02\pm0.06\ M_\odot$, thereby confirming the black hole nature of
the compact object.

In addition to orders \#10--14, and as discussed in the following
section, we also performed a radial velocity analysis of order \#9 for
the purpose of measuring the mass ratio $q$ and investigating the
effects of disk veiling.  The best-fit parameters for order \#9 are
included in Table~\ref{rvc_table}.  However, we do not include the results
for this order in our final determination of the radial velocity
parameters because relatively few spectra generated useful velocity data
(39 out of 72) and because the quality of the fit is relatively poor
($\chi^2/\nu=1.55$).

\begin{center}
\begin{deluxetable}{lcl}
\tabletypesize{\footnotesize}
\tablecaption{Orbital Parameters for \nmus\ \label{orbital_table}}
\tablewidth{0pt}
\tablehead{ \colhead{Parameter} & \colhead{} & \colhead{Value}
}
\startdata
Orbital period $P$ (days) & & $0.43260249\pm0.00000009$ \\
$K_2$ velocity (km s$^{-1}$) & & $\ \ \ \ \ \ \ \ \ \ 406.8\pm2.7$ \\
$\gamma$ velocity (km s$^{-1}$) & & $\ \ \ \ \ \ \ \ \ \ \ \ 14.2\pm6.3$ \\
$T_0$, spectroscopic (HJD$-$2454900) & & $\ \ \ \ 46.90278\pm0.00062$ \\
Mass function $f(M)$ ($M_\odot$) & & $\ \ \ \ \ \ \ \ \ \ \ \
3.02\pm0.06$ \\
$v\sin i$ (km s$^{-1}$) & & $\ \ \ \ \ \ \ \ \ \ \ \ 85.0\pm2.6$ \\
Mass ratio $q$ & & $\ \ \ \ \ \ \ \ \ \ 
0.079\pm0.007$
\enddata
\tablecomments{The quoted uncertainties are at the $1\sigma$ level of
  confidence and include an allowance for systematic error; see
  \S\ref{discuss:sys}.}
\end{deluxetable}
\end{center}


\section{Rotational Velocity and Disk Veiling Measurements}\label{vsini}

\subsection{Rotational Broadening and Mass Ratio}\label{vsini:vsini}

As the next step toward measuring the mass of the black hole, we obtain
a precise estimate of the mass ratio $q$. For a short-period,
mass-exchange binary like \nmus, one can quite confidently assume that
the system is tidally-locked and that the secondary fills its Roche
lobe (Wade \& Horne 1998).  In this case, $q$ can be simply determined
by measuring the rotational broadening of the stellar photospheric
lines.  Specifically, one measures the radial component of the
rotational velocity of the secondary star, \vsini\, and uses the
following equation (Wade \& Horne 1998),
\begin{eqnarray}
  \frac{v \sin i}{K_2} = 0.462q^{1/3}(1+q)^{2/3}.
\end{eqnarray}

\begin{figure*}[t]
    \centering
    \includegraphics[width=6in]{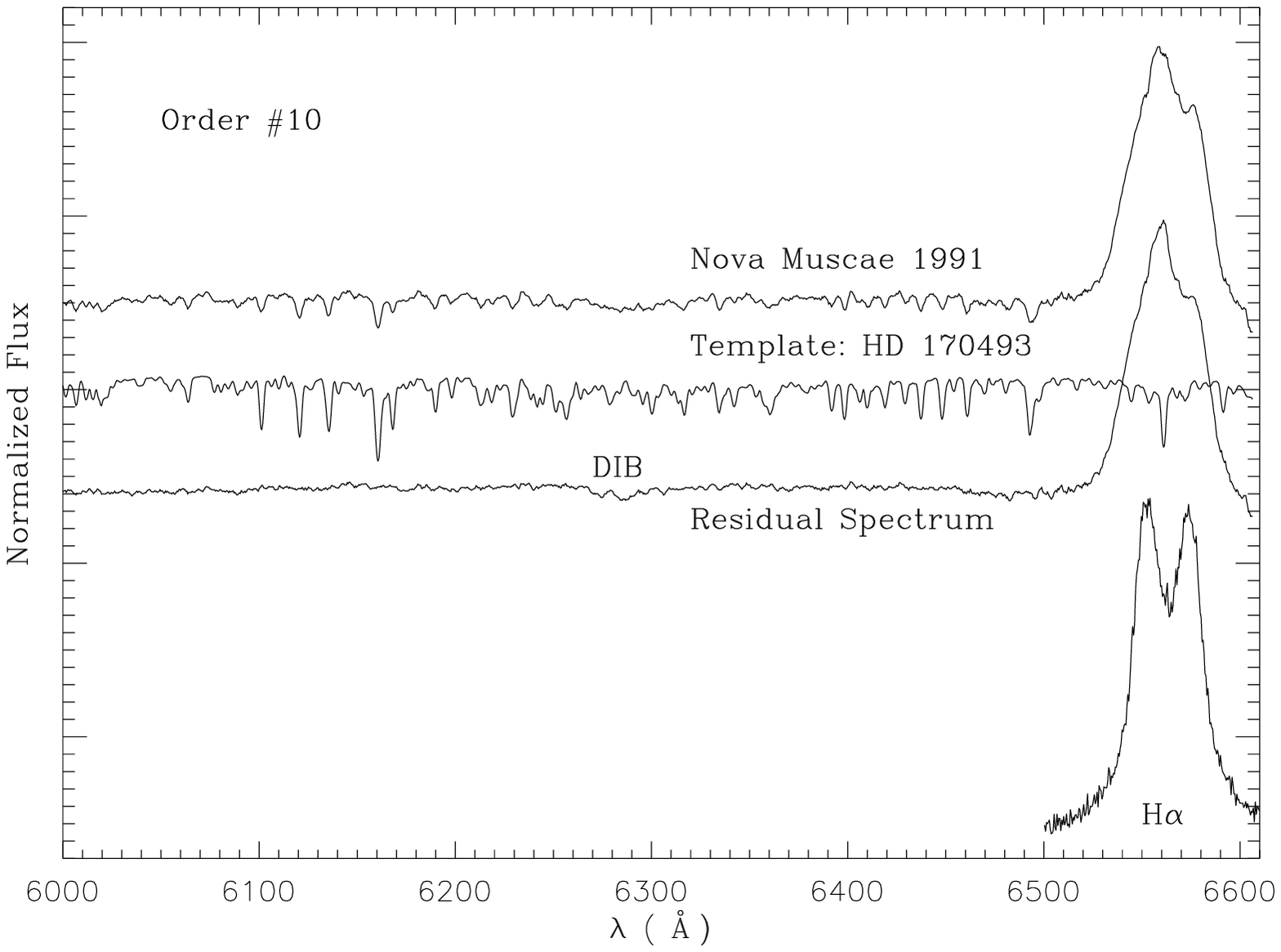}
    \includegraphics[width=6in]{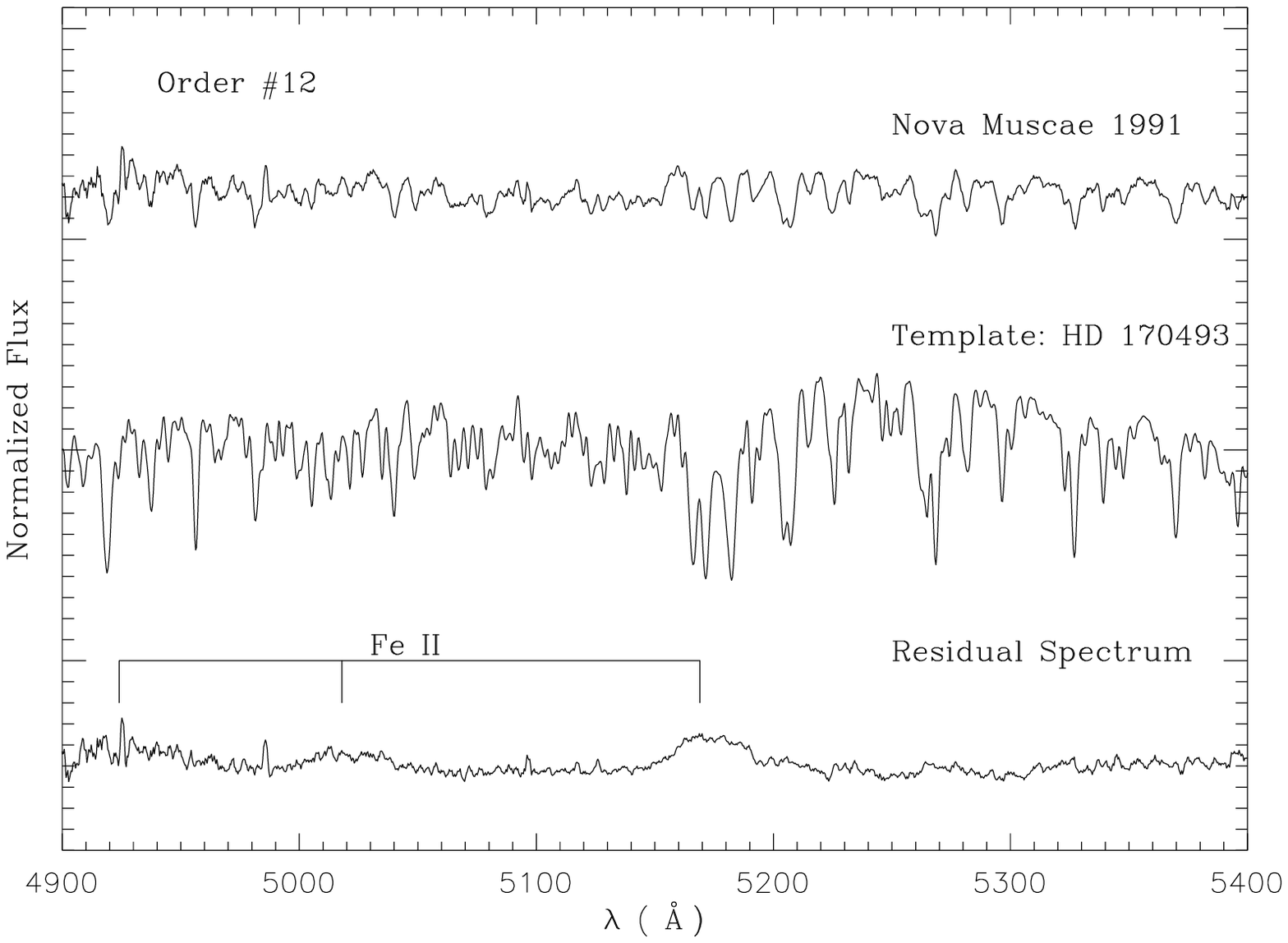}
    \caption{\footnotesize{The results of applying the optimal
        subtraction procedure for order \#10 (upper panel) and order
        \#12 (lower panel). Each panel contains the average normalized
        spectrum of \nmus, the normalized spectrum of the template star
        HD~170493, and the residual spectrum after optimal
        subtraction. The \ha\ emission line and strong DIB feature are
        labeled for order \#10. The upper panel also contains the
        averaged H$\alpha$ emission line in the observer's frame, which
        shows the characteristic double-peaked profile. The Fe~{\sc ii}
        multiplet 42 features in the residual spectrum of order \#12 are
        labeled in the lower panel. All the spectra are smoothed with a
        5-pixel boxcar for the purpose of presentation.}
  \label{resid_fig}}
\end{figure*}%

\begin{figure*}[t]
    \centering
    \includegraphics[width=4.5in, angle=90]{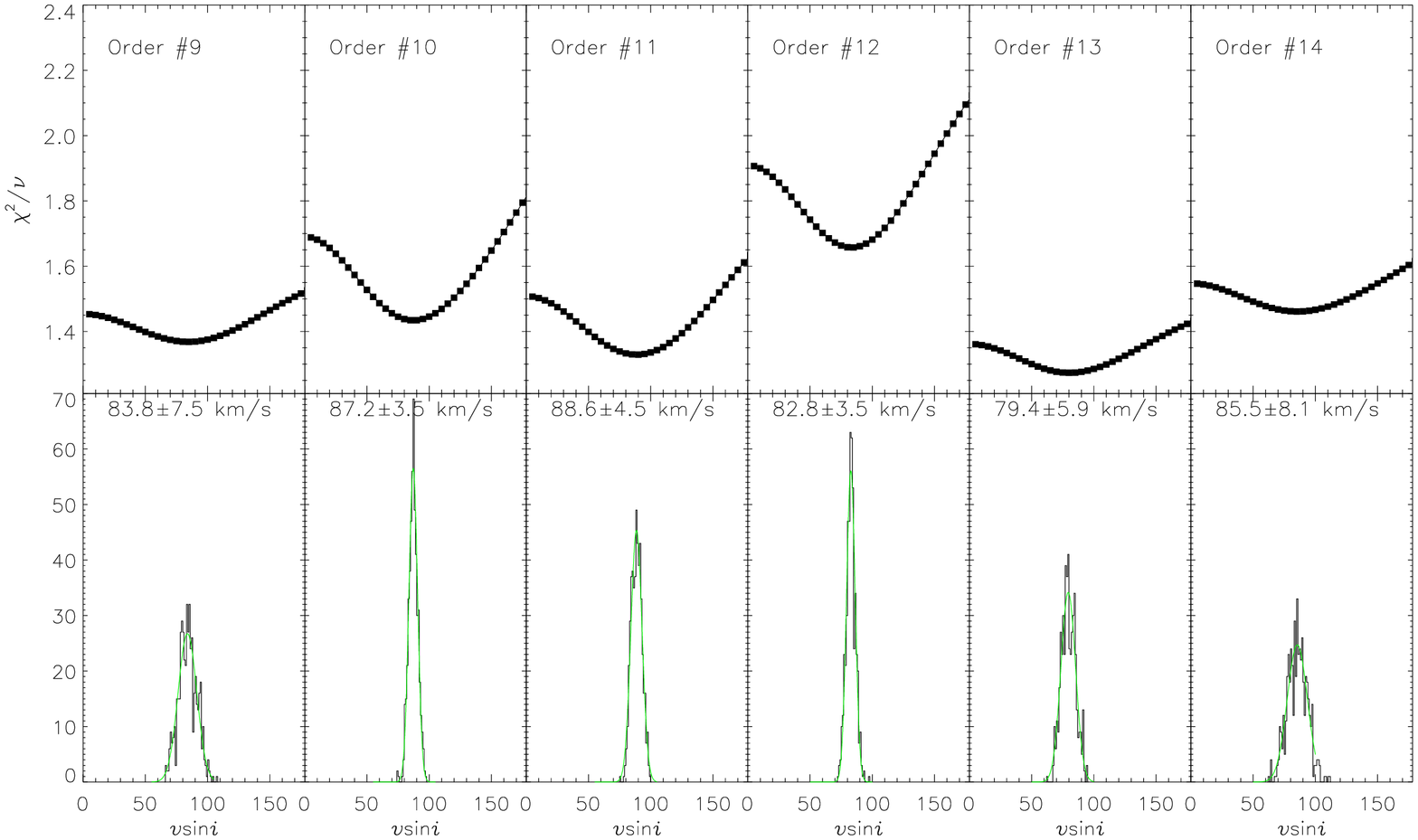}
    \caption{\footnotesize{Measurement of the rotational velocity
    \vsini\ in MagE orders \#9--14. For each order, the upper panel
    shows the curve of \vsini\ vs. $\chi^2$, which we fitted with a cubic
    polynomial. Our adopted value of \vsini\ is the value corresponding
    to the minimum in $\chi^2$.  The lower panel illustrates the method
    we used to estimate the uncertainty in \vsini\ via the bootstrap
    method described in the text.  The histogram of \vsini\ is fitted
    with a Gaussian in determining the uncertainty.}
    \label{vsini_fig}}
\end{figure*}%

In measuring \vsini\, we use the method of {\it optimal subtraction}.
Fig.~\ref{resid_fig} illustrates the principle of the method using the
data for orders \#10 and \#12. Before doing the optimal subtraction, we
first normalized both the object and template spectra using a 5th-order
polynomial.  The spectra of \nmus\ contains contributions from both the
secondary (numerous rotationally-broadened absorption features)
and the accretion disk (broad disk emission lines like \ha).  The
best-match template spectrum will have virtually the same set of
absorption features as NovaMus; however, the features in the object
spectrum will be shallower because they are diluted by the disk
continuum emission.  In obtaining the best match of a given template
spectrum to the object spectrum, one repetitively performs a pair of
operations. First, one broadens the narrow-line template spectrum, and
then multiplies the spectrum by the factor $f_{\rm star}$ ($0\leqslant
f_{\rm star}\leqslant 1$), which represents the fraction of the light
contributed by the secondary. Finally, one subtracts the trial template
spectrum from the spectrum of \nmus\ and determines quantitatively (see
below) the template spectrum that is freest of photospheric absorption
features. The results of such an analysis for two orders are shown in
Fig.~\ref{resid_fig}.  The residual spectrum for order \#10 (upper panel)
only shows the broad H$\alpha$ disk line and the DIB at
$\lambda$6280. The only clear spectral feature in the residual spectrum
of order \#12 (Fig.~\ref{resid_fig}; lower panel) is the Fe~{\sc ii}
multiplet 42 at $\lambda$4924, $\lambda$5018, and $\lambda$5169 (Moore
1972), a feature that is also present in the spectra of A0620$-$00
(Marsh et~al. 1994) and the neutron-star transient EXO~0748$-$676 (Ratti
et~al. 2012).

In practice, we use the spectrum of our velocity-template star HD~170493
and the phase-averaged spectra of \nmus\ for each order to increase the
$S/N$.  Before averaging the spectra for a given order, we shifted the
individual spectra to the rest frame of HD~170493 using the data in
Table~\ref{rvc_table}.  We first broadened the template spectrum using a
grid of trial values from 5 to 300~km~s$^{-1}$ in steps of
5~km~s$^{-1}$.  We also artificially smeared the template spectrum using
the appropriate phase and integration time in order to simulate the
velocity smearing during an observation of \nmus.  A $\chi^2$ test was
performed on the residual spectrum to determine a minimum and the
corresponding approximate value of \vsini. Now, starting with this
approximate value, we repeated the procedure with a trial grid extending
from 60 to 120~km~s$^{-1}$ in steps of 1~km~s$^{-1}$.  Making use of the
\verb+rbroad+ and \verb+optsub+ tasks in the {\sc molly}, curves of
\vsini\ vs. $\chi^2$ were generated (see Fig.~\ref{vsini_fig}) and
fitted with a 3rd-order polynomial; our adopted value of \vsini\
corresponds to the minimum of the fitted curve.  All the emission lines,
interstellar features, and telluric features are masked out during the
optimal subtraction. 

The effect of limb darkening needs to be properly considered in the
optimal subtraction.  Since this effect varies with wavelength, we
calculated a linear limb darkening coefficient for each order of the
MagE spectra based on the data given by Claret et~al. (2012).  We
retrieved from their table the limb darkening coefficients for the
$UBVRI$ bands.  For the effective temperature of the companion star, we
adopted the value $T_{\rm eff}=4400$~K.  The star's surface gravity
$\log g$ (in cm$^{-2}$~s$^{-1}$) is securely in the range 4.0--4.5, and
for each band we adopted the average of the tabulated values of the limb
darkening coefficient for $\log g=4.0$ and $\log g=4.5$.  Finally we
linearly interpolated between the $UBVRI$ bands using the central
wavelength of each order, thereby obtaining the values we adopt for the
limb darkening coefficients (see the third column in
Table~\ref{vsini_table}), which range from 0.71 (Order \#9) to 0.91
(Order \#14).

\begin{center}
\begin{deluxetable}{ccccc}
\tabletypesize{\footnotesize}
\tablecaption{Limb Darkening Coefficient and Rotational Velocity\label{vsini_table}}
\tablewidth{0pt}
\tablehead{ \colhead{Order \#} & \colhead{Central $\lambda$ (\AA)} &
  \colhead{u (limb darkening)} & \colhead{\vsini\ (km~s$^{-1}$)} &
    \colhead{Disk Veiling (\%)}  
}
\startdata
9 & 6800 & 0.706 & $83.8\pm7.5$ & $40.3\pm3.4$ \\
10 & 6130 & 0.762 & $87.2\pm3.5$ & $51.4\pm1.3$ \\
11 & 5590 & 0.808 & $88.6\pm4.5$ & $56.7\pm1.4$ \\
12 & 5140 & 0.845 & $82.8\pm3.5$ & $59.3\pm1.0$ \\
13 & 4750 & 0.878 & $79.4\pm5.9$ & $55.8\pm2.1$ \\
14 & 4400 & 0.908 & $85.5\pm7.1$ & $69.1\pm1.5$ \\
\hline
Average & & & $85.0\pm1.3$ &
\enddata
\tablecomments{The quoted uncertainties are at the $1\sigma$ level of
  confidence.}
\end{deluxetable}
\end{center}

We utilized the bootstrap method to determine the uncertainty in \vsini\
following Steeghs \& Jonker (2007) and Ratti et~al. (2013). We generated
500 simulated \nmus\ spectra using the {\sc molly} task \verb+boot+. The
bootstrap method randomly selects data points from the original spectrum
while keeping the same total number of data points in the spectrum. For
each simulated spectrum, \vsini\ was measured following the same
procedure as described above. The 500 \vsini\ values follow a Gaussian
distribution (see Fig.~\ref{vsini_fig}). The mean of the Gaussian
distribution is very close (within 1~km~s$^{-1}$) to the fitted values
of \vsini\ discussed above. For each order, we take the mean of the
Gaussian distribution as the final measurement of \vsini\, and the
standard deviation of the mean $\sigma$ as the uncertainty.  The
measured values of \vsini\ for six orders (\#9--14) are presented in 
Fig.~\ref{vsini_fig} and Table~\ref{vsini_table}).

The error-weighted mean \vsini\ of all six orders is $\langle v \sin i
\rangle = 85.0\pm1.3$~km~s$^{-1}$. After including an allowance for
systematic uncertainty, related to the choice of template spectrum and
the limb darkening coefficient (see \S\ref{discuss:sys}), our final
adopted value is \vsini = $85.0\pm2.6$~km~s$^{-1}$. Using this value and
our $K$-velocity for the secondary ($K_2=406.8\pm2.7$~km~s$^{-1}$;
\S\ref{rvc:fit}), and Eqn.\ (3), we obtain for the mass ratio the value
$q=0.079\pm0.007$.  Using this mass ratio and the value of the mass
function given in \S\ref{rvc:fit}, the masses of the black hole and
secondary star are $M=3.52\pm0.07(\sin^{-3} i)\ M_\odot$ and
$M_2=0.28\pm0.03(\sin^{-3} i)\ M_\odot$, respectively.

\begin{figure}[t]
    \centering
    \includegraphics[width=3.7in]{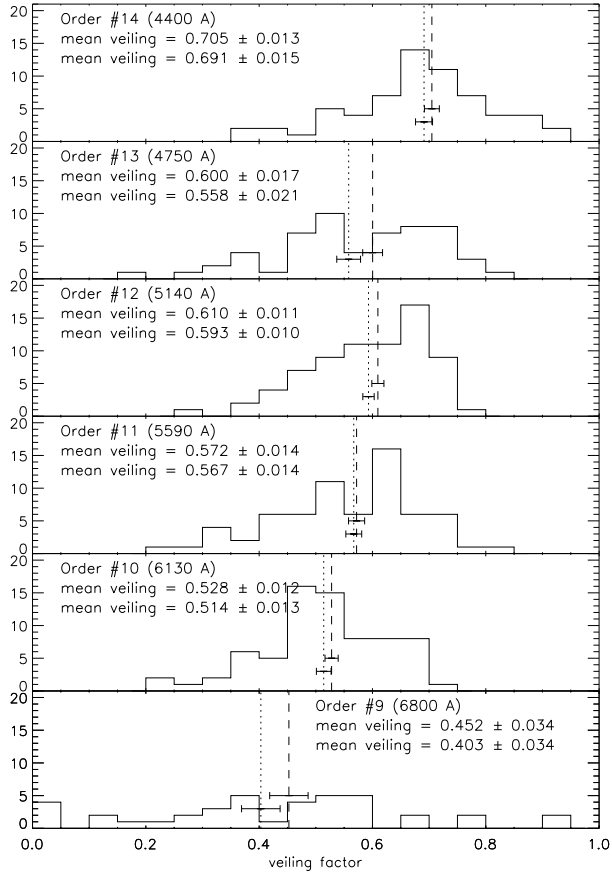}
    \caption{\footnotesize{Measurement of the disk veiling fraction in
    order \#9--14. In each panel, the dotted line indicates the disk
    veiling fraction (labeled by the ``mean veiling 1'' value) of the 
    averaged spectrum. The histogram shows our disk veiling
    measurements for individual spectra; the dashed line
    indicates their mean value as obtained from the histogram (labeled by the
    ``mean veiling 2'' value).}
    \label{veil_fig}}
\end{figure}%

\subsection{Disk Veiling}\label{vsini:veil}

The disk veiling factor $f_{\rm disk}$, i.e., the fraction of the total
light that is non-stellar, is also determined by the optimal subtraction
method.  For each order, we first broaden the HD 170493 template
spectrum using the values of \vsini\ determined in \S\ref{vsini:vsini},
and then optimally subtract it from the spectrum of \nmus.  The procedure
gives the fraction of light due to the secondary star, $f_{\rm star}$,
for the template spectrum that corresponds to the minimum value of
$\chi^2$.  The disk veiling factor is then simply $f_{\rm disk} = 1 -
f_{\rm star}$.  As in determining \vsini\, we computed $f_{\rm disk}$
using for each order the averaged, rest-frame spectrum of \nmus. The
values we adopt for the disk veiling factor for the six orders are
indicated in Fig.~\ref{veil_fig} by dotted lines and are listed in the
last column of Table~\ref{vsini_table}.   

As a check on the robustness of our measured values of $f_{\rm disk}$,
we repeated the optimal subtraction procedure for each spectrum
individually, using only those spectra that produced a reliable
measurement of radial velocity (see \S\ref{rvc}).  A histogram of
$f_{\rm disk}$ for each order is shown in Fig.~\ref{veil_fig}; the
dashed line in each panel represents the weighted mean value of the disk
veiling factor.  These alternative measurements of $f_{\rm disk}$, 
are consistent within $\sim
1\sigma$ (for all orders) with the values determined as described above
using the phase-averaged spectra.  We note that for five of the six
orders $f_{\rm disk}$ is greater than 0.5, i.e., the disk is brighter
than the secondary star.

\begin{figure}[t]
    \centering
    \includegraphics[width=3.5in]{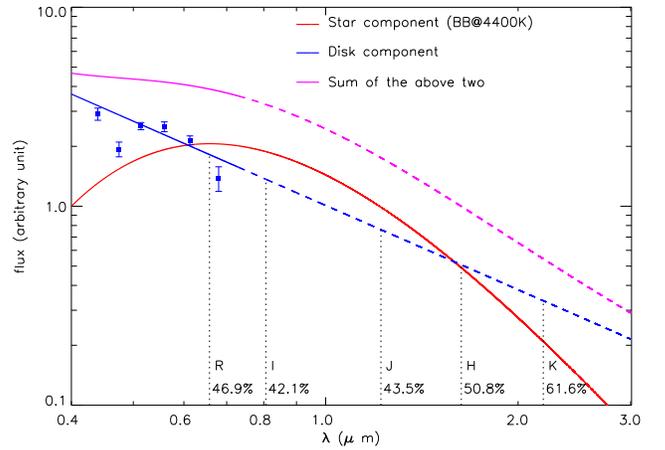}
    \caption{\footnotesize{The simplistic spectral energy distribution
        of \nmus, decomposed into the secondary star component (red line)
        and the accretion disk component (blue line). The pink line
        shows the sum of the two components. The solid blue and pink
        lines cover the observed wavelength range for MagE orders
        \#9--14. The dashed lines are extrapolations to longer
        wavelengths. The star component is represented by a 4400~K
        blackbody spectrum. The blue line represents a power-law model
        obtained by fitting the observed relative flux of disk emission
        (filled blue squares) in the six orders, excluding the data
        point for order \#13 (the second blue square from the left),,
        which is a clear outlier. The black dotted lines show the
        effective wavelengths of the $R$, $I$, $J$, $H$, $K$ bands. Our
        estimate of the disk veiling factor in each band is also
        labeled. }
    \label{starmodel_fig}}
\end{figure}

Fig.~\ref{veil_fig} indicates that the disk veiling factor generally
decreases with increasing wavelength, i.e., the disk veiling is less
significant for redder wavebands.  However, this does not necessarily
mean that disk veiling can be ignored, even in the $J$ and $K$ bands, as
claimed by Gelino et~al. (2001) and others.  We utilize a simplistic
spectral energy distribution (SED) model and an extrapolation to
illustrate this point (see Fig.~\ref{starmodel_fig}).  Our crude SED by
no means represents an accurate model of the disk emission.  For the
stellar component, we assumed blackbody emission at 4400~K (red line in
Fig.~\ref{starmodel_fig}; see \S\ref{rvc} and Fig.~\ref{temp_fig}), and
we then computed the relative disk flux in each order (filled blue
squares) based on our measurements of $f_{\rm disk}$.  We fitted these
estimates of disk flux using a power-law model to represent the SED of
the disk component and then extrapolated the power law into the
near-infrared (dashed blue line).  

At shorter wavelengths, the disk component does indeed decrease rapidly
with increasing wavelength so that the stellar component becomes
dominant as one approaches the $R$-band.  However, at longer wavelengths
the blackbody component begins to fall rapidly, and disk veiling reaches
a minimum ($\approx40\%$) at around $1\ \mu$m and then begins to
increase.  For our data and this crude model, we estimate that $f_{\rm
  disk}$ for the $I$, $J$, $H$, $K$ bands is in the range of 40--60\%
(see the black dotted lines in Fig.~\ref{starmodel_fig}), which is
consistent with the estimates of Reynolds et~al.\ (2008) for \nmus\ of
40--50\% in the near-infrared.  We note that $f_{\rm disk}$ in the
near-infrared could be lower than our estimate if the disk component of
emission has a thermal spectrum that falls below the extrapolated
power-law spectrum, which we assume.  However, we conclude that it is not
warranted to assume that the effects of veiling are negligible in the
near-infrared.

\begin{figure*}[t]
    \centering
    \includegraphics[width=7.0in]{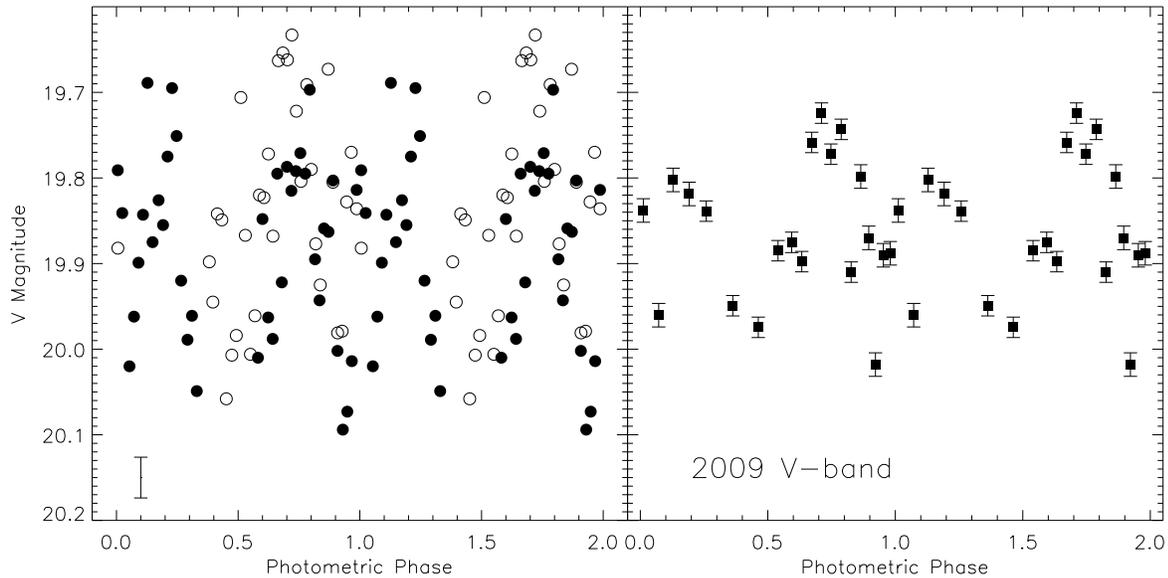}
    \caption{\footnotesize{The original (left panel) and phase-binned 
    (right panel) V-band light curves of \nmus, acquired with the 
    du~Pont 2.5-m Telescope. In the left panel, the open (filled) circles 
    are the photometry data points obtained during the first (second) 
    night of observation. The error bar in the lower left corner represents 
    the typical uncertainty. In the right panel, the 
    light curve is binned into 20 phase bins, each of which has three or 
    four individual data points.}
    \label{lc_fig}}
\end{figure*}%

\section{Light Curve}\label{lc}

Our calibrated and phase-folded $V$-band light curve, which was obtained
in exposures that obtained simultaneously with the spectroscopic
exposures (see \S\ref{obs:photo}), is shown in Fig.~\ref{lc_fig}. Zero
photometric phase, which is equivalent to 0.75 spectroscopic phase,
corresponds to the inferior conjunction of the secondary.  Two versions
of the light curve are shown, unbinned and binned into 20 phase bins,
each with 3--4 data points per bin.

During our observations, \nmus\ was $\sim 0.5$ magnitude brighter on
average in the $V$-band than it was in earlier photometric monitoring
campaigns (e.g., during 1992--1995, as reported by Orosz et~al. 1996).
Our $V$-band light curve shows substantial aperiodic flickering (see
\S\ref{discuss:comp}).  Nevertheless, the phase-binned light curve
exhibits the characteristic signatures of ellipsoidal modulation, with
two minima at phases $\sim0.0$ and $\sim0.5$ and two maxima at phases
$\sim0.25$ and $\sim0.75$. The two maxima do not have equal amplitude as
one would expect for an ideal ellipsoidal modulation, which could be
caused by a hot spot in the accretion disk. Such unequal maxima have
frequently been observed in the light curves of other black hole
binaries in quiescence, e.g., A0620$-$00 (McClintock \& Remillard 1986;
Cantrell et al. 2010).


\section{Discussion}\label{discuss}

\subsection{Assessment of Systematic Uncertainties}\label{discuss:sys}

We identify sources of systematic error, and obtain estimates of their
magnitude, first for the orbital parameters $K_2$ and $T_{\rm 0}$ and
the systemic velocity $\gamma$, and then for the rotational velocity
\vsini.

\begin{center}
\begin{deluxetable*}{lcccccccc}
\tabletypesize{\footnotesize}
\tablecaption{Radial Velocity Analysis Results for Different Templates (Order \#12)\label{rvc_table2}}
\tablewidth{0pt}
\tablehead{ \colhead{Templates} & \colhead{$B-V$} & \colhead{Template $\gamma$} & \colhead{TDR} & \colhead{$K_2$} &
  \colhead{$\gamma$} & \colhead{$T_0-2454900$} & \colhead{No. of} &
  \colhead{$\chi^2/\nu$}  \\
\colhead{(Order \#12)} & & \colhead{(km~s$^{-1}$)} & & \colhead{(km~s$^{-1}$)} & 
  \colhead{ (km~s$^{-1}$)} & \colhead{(d)} & \colhead{Object Spectra} &
  \colhead{}  
}
\startdata
HD~147776 & 0.95 & $7.4\pm0.1$ & $60.8$ & $402.0\pm2.7$ & $17.1\pm2.0$ & $46.90339\pm0.00044$ & $72$ &
$0.83$ \\
HD~130992 & 1.02 & $-57.1\pm0.1$ & $64.1$ & $401.9\pm2.7$ & $10.9\pm2.0$ & $46.90333\pm0.00044$ & $72$ &
$0.79$ \\
HD~31560 & 1.09 & $6.4\pm0.1$ & $74.3$ & $401.6\pm2.7$ & $11.4\pm2.0$ & $46.90332\pm0.00044$ & $72$ &
$0.81$ \\
\hline\\
HD~170493 & 1.11 & $-55.1\pm0.1$ & $78.4$ & $401.2\pm2.6$ &
$9.7\pm1.9$ & $46.90327\pm0.00044$ & $72$ &
$0.87$ \\ 
HD~170493 (broadened) & 1.11 & $-55.1\pm0.1$ & $76.7$ & $401.7\pm3.0$ &
$10.5\pm2.2$ & $46.90347\pm0.00050$ & $72$ &
$0.60$ \\ \\ 
\hline
HD~131977 & 1.11 & $26.8\pm0.1$ & $65.9$ & $401.6\pm2.7$ & $10.9\pm2.0$ & $46.90326\pm0.00045$ & $72$ &
$0.79$ \\
HD~156026 & 1.16 & $0.1\pm0.1$ & $61.6$ & $401.9\pm2.8$ & $4.6\pm2.0$ & $46.90344\pm0.00046$ & $72$ &
$0.87$ \\
HD~120467 & 1.26 & $-37.8\pm0.1$ & $63.1$ & $401.9\pm2.8$ & $5.5\pm2.1$ & $46.90365\pm0.00047$ & $72$ &
$0.79$ \\
\enddata
\end{deluxetable*}
\end{center}

\subsubsection{Orbital Parameters and Systemic Velocity}

We first consider the uncertainty associated with the choice of template
spectrum.  While we chose the spectrum of HD~107493 because it gave the
maximum TDR value (\S\ref{rvc:temp}), obviously this spectrum is not
identical to that of the Roche-lobe-filling companion star.  Focusing on
the high-quality data for order \#12, we repeated our analysis for six
other template spectra with TDR values $>60$ (see Fig.~\ref{temp_fig});
the results are summarized in Table~\ref{rvc_table2}.  The spread in the
values of $K_2$ is $<0.8$~km~s$^{-1}$, which is significantly smaller
than the statistical uncertainty.  As our estimate of the systematic
error in $K_2$ associated with the choice of the template, we take the
standard deviation of the seven values of $K_2$, which is
0.3~km~s$^{-1}$.  In like manner, we obtain estimates of systematic
error of 4.2~km~s$^{-1}$ for $\gamma$, and of 0.00013~day for $T_0$. 

A second source of systematic uncertainty is the effect of the different
widths of the photospheric lines for \nmus\, which are rotationally
broadened, and for the slowly-rotating template star, which are narrow.
To assess this effect, we broadened the lines of our chosen template
(HD~107493) to 85~km~s$^{-1}$ (\S\ref{vsini:vsini}) and repeated the
radial velocity analysis.  The change in $K_2$ is 0.5~km~s$^{-1}$
(Table~\ref{rvc_table2}), which we adopt as our estimate of systematic
error; we obtain similar estimates of systematic error for $\gamma$
and $T_0$ of 0.8~km~s$^{-1}$ and 0.00020 day, respectively. 

For each of the three parameters in question, we obtain our final
estimate of error, which is given in Table 2, by linearly adding the
larger of the two systematic errors discussed above onto the statistical
error obtained in \S\ref{rvc:fit} (given in Table~1). 

\begin{center}
\begin{deluxetable}{lcccc}
\tabletypesize{\footnotesize}
\tablecaption{Rotational Velocity for Different Templates (Order \#12)\label{vsini_table2}}
\tablewidth{0pt}
\tablehead{ \colhead{Template} & \colhead{$B-V$} & \colhead{u (limb darkening)} & \colhead{\vsini\ (km~s$^{-1}$)} 
}
\startdata
HD~170493 & 1.11 & 0.845 & $82.8\pm3.5$\\
\hline
HD~170493 & 1.11 & 0.745 & $81.7\pm3.6$ \\
HD~170493 & 1.11 & 0.945 & $84.3\pm3.7$ \\
\hline
HD~147776 & 0.95 & 0.823 & $82.9\pm4.6$ \\
HD~130992 & 1.02 & 0.842 & $82.3\pm4.5$ \\
HD~31560  & 1.09 & 0.845 & $82.4\pm4.2$ \\
HD~131977 & 1.11 & 0.845 & $82.3\pm4.0$ \\
HD~156026 & 1.16 & 0.835 & $83.8\pm4.3$ \\
HD~120467 & 1.26 & 0.796 & $85.5\pm5.2$ \\
\enddata
\tablecomments{The quoted uncertainties are at the $1\sigma$ level of
  confidence.}
\end{deluxetable}
\end{center}

\subsubsection{Rotational Velocity}

We identify two sources of systematic uncertainty for \vsini, namely,
the value adopted for the limb darkening coefficient and the choice of
template spectrum.  Again, using the data for order \#12, we considered
an uncertainty of $\pm0.1$ in our adopted value of 0.845 (Table~3) and
find that this corresponds to an uncertainty of $\le1.5$~km~s$^{-1}$
(see Table~\ref{vsini_table2}).  We then compared values of \vsini\
derived using once again the six template spectra with TDR values $>60$.
For each template spectrum, we used the appropriate limb darkening
coefficient (Claret et~al. 2012) based on the $B-V$ color and
corresponding effective temperature of the star.  The standard deviation
for the six values of \vsini\ is 1.2~km~s$^{-1}$.  Using these two
estimates of systematic error as a guide, we allow for systematic error
by simply doubling the statistical error bar obtained in
\S\ref{vsini:vsini} (1.3~km~s$^{-1}$).  Our final adopted value of
\vsini\, given in Table~2, is therefore $85.0\pm2.6$~km~s$^{-1}$

\subsection{Comparison to Previous Work}\label{discuss:comp}

We have obtained precise measurements of the mass function $f(M)$ and
mass ratio $q$ for \nmus\ using multi-order echellette spectral data
obtained with Magellan/MagE. Compared to the data analyzed in previous
work (e.g., Orosz et~al. 1996 and Casares et~al. 1997), our spectra have
higher spectral resolution (by a factor of $\sim3$--5) and provide more
complete phase coverage.  We also observed many more template stars
(38), all of them during the same observing run, which has enabled us to
choose a template spectrum that is an exceptionally close match to the
spectrum of \nmus.  Analyzing data independently for several orders is
also a feature of our work.  All of these beneficial factors have
contributed to quality of our work.

Our value for the radial velocity amplitude
($K_2=406.8\pm2.7$~km~s$^{-1}$) is very close to that obtained by
Orosz et~al. (1996; $K_2=406\pm7$~km~s$^{-1}$), while the uncertainty
has been reduced significantly. Meanwhile, our $K_2$ value is barely
consistent with that of Casares et~al. (1997;
$K_2=420.8\pm6.3$~km~s$^{-1}$), differing by $\approx2\sigma$.  We
have substantially increased the precision of the orbital period (by a
factor of $>30$) by extending the baseline of observations from those
made earlier (Remillard et al.\ 1992; Orosz et al.\ 1996) out to our
observations in 2009. 

Our mass ratio ($q=0.079\pm0.007$) agrees with the approximate value
($0.0476<q<0.0833$) found by Antokhina \& Cherepashchuk (1993), which
was obtained by modeling an $I$-band light curve.  It is significantly
smaller than the less reliable estimate of $q=0.133\pm0.019$ obtained by
Orosz et~al. (1996; see \S\ref{intro}).  Casares et~al. (1997), who used
the same direct method of measurement we used, likewise obtained a mass
ratio greater than ours, $q=0.13\pm0.04$.  (Their estimate of the
rotational broadening is $v \sin i =106\pm13$~km~s$^{-1}$ compared to
our value of $v \sin i =85.0\pm2.6$~km~s$^{-1}$.)  The relative
virtues of our measurement is our better spectral resolution
($\approx60$~km~s$^{-1}$ at $\sim5000$~\AA) and our ample selection of
template spectra, which allowed us to closely match the spectrum of
the donor star; Casares et al.\ had applied a single template spectrum
of a K0V star. 

An important feature of our work is the simultaneity of the
spectroscopic and photometric observations, which allows one to
accurately correct for the effects of disk veiling, despite the
aperiodic variability of the non-stellar component of light.
Unfortunately, during our observation \nmus\ was in an ``active''
quiescent state (Cantrell et~al. 2008), and the light curve is strongly
affected by aperiodic flickering and does not provide a useful
constraint on the inclination angle.  Nevertheless, our simultaneous
observations provide very secure measurements of the disk veiling factor
at six wavelengths.  In a subsequent paper, we will use these data in
conjunction with light curves available in the literature, which were
obtained in the ``passive'' quiescent state (Cantrell et~al. 2008) to
constrain the inclination angle and black hole mass.

\subsection{The Li $\lambda$6708 Feature}\label{discuss:li}

\begin{figure*}[t]
    \centering
    \includegraphics[width=6.0in]{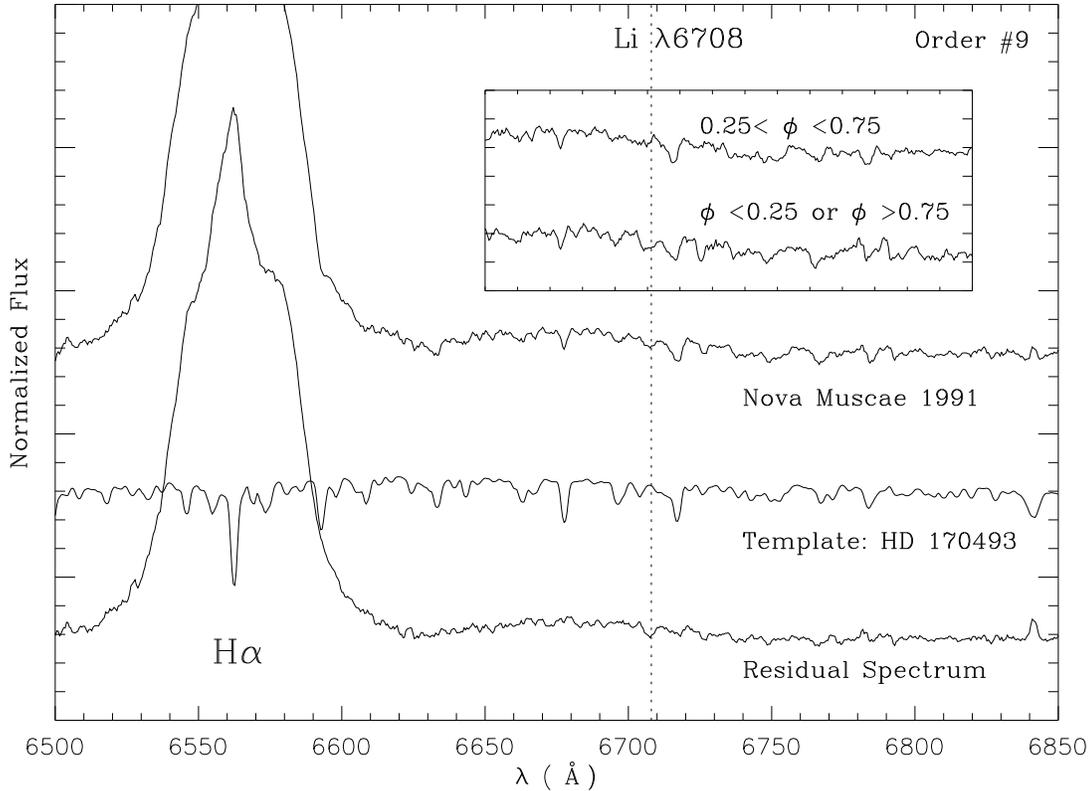}
    \caption{\footnotesize{The averaged spectrum of \nmus\ (top), the 
    template spectrum of HD~170493 (middle), and the residual spectrum 
    (bottom) after optimal subtraction, all for MagE order \#9 and in 
    the heliocentric frame. The 
    location of Li $\lambda$6708 is indicated by the dotted line. This Li 
    absorption feature is marginally detected ($\sim3.5\sigma$). The
    inset plot shows the phase-resolved \nmus\ spectra around Li
    $\lambda$6708. The Li feature is detected in the lower
    spectrum, in which the photometric phase $\phi$ is around zero
    ($\phi>0.75$ or $\phi<0.25$). There is no clear Li feature in the
    upper spectrum around photometric phase 0.5 ($0.25<\phi<0.75$). 
    All the spectra are smoothed with a 
    5-pixel boxcar for the purpose of presentation.}
    \label{li_fig}}
\end{figure*}%

The Li $\lambda$6708 absorption line was previously detected in the
spectra of \nmus\ (Mart{\'{\i}}n et~al. 1996). This feature also appears
in the spectra of other similar black hole soft \xray\ transients like
A0620$-$00 (Marsh et~al. 1994) and GS~2000$+$25 (Filippenko
et~al. 1995).  This feature is potentially relevant in exploring
the Li production mechanism around a black hole (Mart{\'{\i}}n et~al. 1994). 
Meanwhile, the variation of the line intensity with orbital phase
may reveal effects of irradiation on the formation of the Li
feature (Mart{\'{\i}}n et~al. 1996). 

We find that the Li $\lambda$6708 line is marginally detected in the
phase-averaged spectrum of \nmus\ and also in the residual spectrum
after optimal subtraction (Fig.~\ref{li_fig}).  We find an equivalent
width of EW(Li)$=210\pm60$~m\AA, after correcting for the disk 
veiling in order \#9, for the \nmus\ spectrum averaged over the full
orbital phase. Following Mart{\'{\i}}n et~al. (1996), we investigated
the variation of the EW(Li) with orbital phase (see the inset plot of
Fig.~\ref{li_fig}). At zero photometric phase ($\phi>0.75$ or
$\phi<0.25$), we obtained EW(Li)$=550\pm90$~m\AA, which is slightly
higher than the value in Mart{\'{\i}}n et~al. (1996;
$420\pm60$~m\AA). The Li $\lambda$6708 feature is not detected at 
photometric phase 0.5 ($0.25<\phi<0.75$), which agrees with the
findings in Mart{\'{\i}}n et~al. (1996). We note that our EW(Li) for
\nmus\ is  slightly higher (within 1$\sigma$ ) than the value 
reported for A0620$-$00 of EW$=160\pm30$~m\AA\ by Marsh et~al. (1994).


\begin{acknowledgments}

We thank the anonymous referee for a careful reading of the manuscript and the 
helpful comments. We thank T. Marsh for developing and sharing the {\sc molly} spectral
analysis software. J.E.M. acknowledges the support of NASA grant NNX11AD08G. 
D.S. acknowledges support from the Science and Technology Facilities Council, 
grant number ST/L000733/1. L.J.G. and L.C.H. acknowledge the support by the 
Chinese Academy of Sciences through
grant No. XDB09000000 (Emergence of Cosmological Structures) from the
Strategic Priority Research Program. L.J.G. acknowledges the support
by National Natural Science Foundation of China (grant No. 11333005)
and by National Astronomical Observatories of China (grant
No. Y234031001). L.C.H acknowledge the support by the National Natural
Science Foundation of China through grant No. 11473002.

Support for the design and construction of the Magellan Echellette Spectrograph
was received from the Observatories of the Carnegie Institution of Washington,
the School of Science of the Massachusetts Institute of Technology, and the
National Science Foundation in the form of a collaborative Major Research
Instrument grant to Carnegie and MIT (AST0215989).

\end{acknowledgments}

{\it Facility:} \facility{Magellan:Clay (Magellan Echellette
  Spectrograph)}, \facility{du~Pont}





\end{document}